\documentclass[fleqn,twoside,twocolumn,nofootinbib]{revtex4} 
\usepackage[sec]{ujp} 
\begin{document}
\title[ESTIMATION OF THE EFFICIENCY OF MATERIAL INJECTION ]
{ESTIMATION OF THE EFFICIENCY OF MATERIAL\\ INJECTION~ INTO~ THE~
REFLEX~ DISCHARGE\\
BY SPUTTERING THE CATHODE MATERIAL}%
\author{Yu.V.~Kovtun}
\affiliation{National Science Center ``Kharkiv Institute of Physics
and Technology'',
\\ Nat. Acad. of Sci. of Ukraine}
\address{1, Akademichna Str., Kharkiv 61108, Ukraine}
\email{Ykovtun@kipt.kharkov.ua}
\author{E.I.~Skibenko}%
\affiliation{National Science Center ``Kharkiv Institute of Physics
and Technology'',
\\ Nat. Acad. of Sci. of Ukraine}%
\address{1, Akademichna Str., Kharkiv 61108, Ukraine}%
\email{Ykovtun@kipt.kharkov.ua}
\author{A.I.~Skibenko}
\affiliation{National Science Center ``Kharkiv Institute of Physics
and Technology'',
\\ Nat. Acad. of Sci. of Ukraine}%
\address{1, Akademichna Str., Kharkiv 61108, Ukraine}%
\email{Ykovtun@kipt.kharkov.ua}
\author{V.B.~Yuferov}
\affiliation{National Science Center ``Kharkiv Institute of Physics
and Technology'',
\\ Nat. Acad. of Sci. of Ukraine}%
\address{1, Akademichna Str., Kharkiv 61108, Ukraine}%
\email{Ykovtun@kipt.kharkov.ua}

\udk{533.915} 

\razd{\secv}
\setcounter{page}{901}%
\maketitle



\begin{abstract}
The processes of injection of a sputtered-and-ionized working material
into the pulsed reflex discharge plasma have been considered at
the initial stage of dense gas-metal plasma formation. A calculation
model has been proposed to estimate the parameters of the sputtering
mechanism for the required working material to be injected into the
discharge. The data obtained are in good accordance with
experimental results.
\end{abstract}

\section{Introduction}

The reflex discharge \cite{1}, also known as the Penning discharge, has a
long-term story of its development and research. Now, it is widely used in
various domains of science and engineering \cite{2,3,4,5,6,7}. For instance,
the study of the sputtering process of various materials into the reflex
discharge plasma \cite{5,6,7} is needed for the determination of a possibility
of their usage as constructional ones in thermonuclear reactors. It is worth
distinguishing the problem of introducing a working substance, i.e. which is to be
separated, into the work space of a magnetoplasma separator on the
basis of reflex discharge \cite{8} or a discharge of any other type. It is
known from the literature \cite{9,10,11,12} that there are a number of
approaches, which can be used for such purposes. Namely, these are the thermal
evaporation \cite{9}, thermal evaporation with subsequent preionization
\cite{10}, thermal evaporation of a substance and its introduction in the form of
supersonic stream \cite{11}, and the sputtering of a separated substance
\cite{12}. In the latter case, plasma is created with the help of a source
functioning on the basis of the electron-cyclotron resonance, and the substance
to be separated is introduced into plasma in the form of an additional wafer
with an applied negative potential. Here, dominating is a scheme (a device)
for introducing the working substance, where the reflex discharge is
applied. In this case, the following sequence of operations is realized.
Plasma is created preliminarily; the corpuscular sputtering of a cathode
material takes place; and the sputtered material penetrates into plasma,
where it undergoes the subsequent ionization. This procedure does not require
additional facilities, as was in work \cite{12}. The previous experience
\cite{8,11,13,14,15} testifies that the reflex discharge is an effective
instrument for producing a multicomponent gas-metal plasma. In this case, the
metal plasma component is formed as a result of the ionization of cathode
material particles that penetrate into the discharge, when the material is
sputtered.

However, the process of cathode material sputtering and its
subsequent ionization in the reflex discharge have not been analyzed
in detail. Especially important is the consideration for impulse
devices, for which the characteristic times of plasma formation are
shorter than the equilibration time of stationary ionization. A
straightforward determination of how much of the heavy fraction of
multicomponent plasma was introduced into the discharge is rather a
difficult task, which requires considerable material expenses.
Therefore, the development of a technique for the quantitative
estimation of the amount of the heavy sputtered plasma component that
penetrates into the discharge is regarded as a useful and necessary
task. Hence, this work aims at analyzing the processes of material
introduction and ionization at the initial stage of producing a dense
gas-metal plasma in a pulsed reflex discharge owing to the
sputtering mechanism. In so doing, we consider it necessary to
choose a calculation model, in the framework of which not only the
processes giving rise to the creation of a gas-metal plasma
(sputtering and ionization) would be taken into account, but also
the processes that are responsible for the uniform filling of
the internal volume with a confining magnetic field -- for instance, it
may have a plug geometry -- with both plasma and the neutral substance
within a finite time interval. In the framework of the proposed
model -- its diagram that illustrates the main stages of
calculation, i.e. sputtering, collisions between sputtered and gas
atoms, ionization, and formation of a gas-metal plasma, is presented
in Fig.~1 -- the following parameters are to be calculated. For the
sputtering stage, these are the dependences of the sputtering
coefficient on the mass and energy of incident ions, as well as on
the angle of incidence, and the total number of sputtered particles.
For the stage of collision between the sputtered atom and the gas
ones, these are the energy spectrum and the average energy of
sputtered atoms, the mean free path of a sputtered atom in the gas,
the diffusion coefficient, and the diffusion time. At the stage of
ionization and gas-metal plasma generation, those parameters include
the time and the degree of ionization of sputtering atoms and the
content of sputtered material atoms in plasma.

\begin{figure}
\includegraphics[width=6cm]{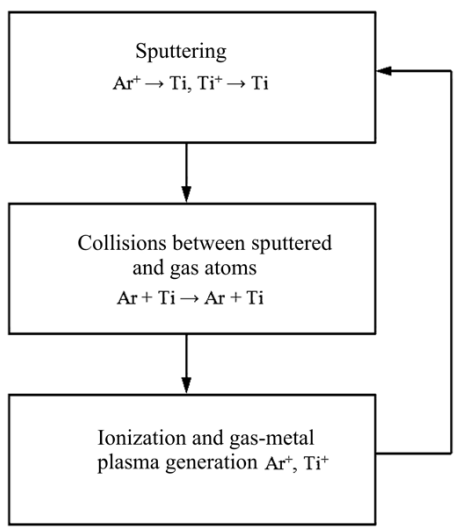}
\vskip-3mm\caption{Diagram of the calculation model  }
\end{figure}

\section{Evaluation of Sputtering Mechanism Parameters}

We will consider the processes of material sputtering and ionization
at the initial stage of dense gas-metal plasma formation in the
pulsed reflex discharge assuming that typical experimental
conditions take place \cite{13,14}: namely, the discharge voltage is
3.5--4~kV, Ar is taken as an ignition gas, cathodes are made of Ti,
and the time interval, during which a density of about $2\times
10^{19}$~m$^{-3}$ is attained, is about 100~$\mu \mathrm{s}$. The
stage of discharge ignition will not be considered now, because such
an analysis was carried out earlier (see, e.g., work \cite{16}).

First, let us consider the processes associated with the interaction between the
plasma and the surface of a solid. A number of processes take place at that
\cite{17}, such as the sputtering, electron emission induced by
the particle-surface interaction, penetration, reflection, and desorption of
stimulated particles, a modification of the near-surface layer, a variation
of the charge state of ions, blistering, and so on. One of the basic processes
leading to the cathode material destruction and, respectively, its
penetration into plasma, is sputtering. The corresponding major
characteristic is the sputtering coefficient $Y$, which depends on the
charge, mass, and energy of bombarding ions, the angle of incidence, the
material that the target is made of, and the target temperature. The
sputtering process has the threshold behavior with respect to the energy. The
dependence of the sputtering coefficient $Y$ on the target material reveals
itself, first, as a function of the mass and the atomic number of target atoms;
second, as a dependence on the surface binding energy of target atoms, $U_{s}
$, which is usually considered to be equal to the sublimation energy per one
atom. For a monoatomic substance, the dependence of the sputtering coefficient
on the ion energy, $Y(E)$, at the normal incidence can be expressed using the
empirical formula \cite{18}
\[
Y(E)=0.042\frac{Q(Z_{2})\alpha ^{\ast }\left( M_{2}/M_{1}\right)
}{U_{s}}\times
\]
\begin{equation}
\times \frac{S_{n}(E)}{1+\Gamma k_{e}\varepsilon ^{0.3}}\left[
1-\sqrt{\frac{E_{th}}{E}}\right] ^{s},
\end{equation}
where the dimensionality of the numerical multiplier is
{\AA}$^{-2}$; $E$ is the energy of an incident ion [eV]; $M_{1}$ and
$M_{2}$ are the masses of an incident ion and a target atom,
respectively [a.m.u]; $E_{\mathrm{th}}$ is the threshold sputtering
energy [eV]; $Q(Z_{2})$ is a dimensionless parameter; $\alpha ^{\ast
}(M_{1}/M_{2})$ is a function (independent of the energy) of the
mass ratio; $S_{n}(E)$ is the nuclear stopping cross-section
[eV$\times \mathrm{\mathring{A}}^{-2}$/atom]; $k_{e}$ is the
Lindhard electronic stopping coefficient; $\varepsilon $ is a
dimensionless energy variable; $s$ is the power exponent, which
weakly depends on the target material; and $\Gamma $ is a factor.
The dependences of the sputtering coefficients for pairs
$\mathrm{Ar}^{+}\rightarrow \mathrm{Ti}$ and
$\mathrm{Ti}^{+}\rightarrow \mathrm{Ti}$ calculated by formula~(1)
are depicted in Fig.~2.

\begin{figure}
\includegraphics[width=\column]{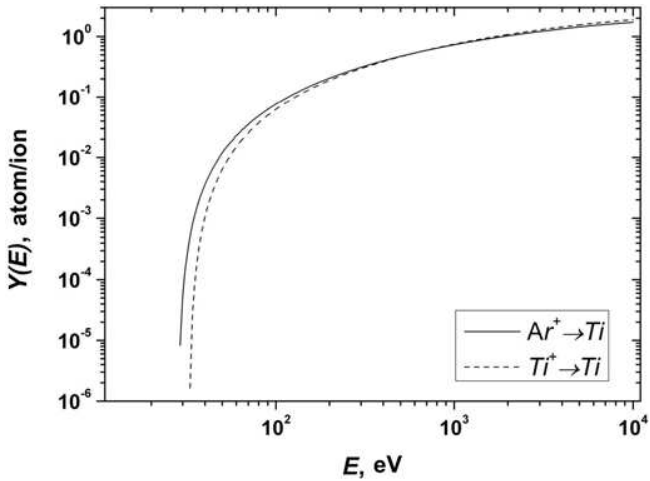}
\vskip-3mm\caption{Dependences of the sputtering coefficient on the
energy of incident ions of various sorts at normal incidence  }
\end{figure}

One can see that the sputtering coefficients for both chosen pairs differ a
little from each other and, at energies of incident ions of 1--4~keV, fall
in the interval 0.73--1.44. The magnitude of threshold sputtering energy
$E_{\mathrm{th}}$ in model \cite{18} amounts to 28 and 32~eV for the pairs
$\mathrm{Ar}^{+}\rightarrow \mathrm{Ti}$ and $\mathrm{Ti}^{+}\rightarrow
\mathrm{Ti}$, respectively. In the reflex discharge, according to the
results of work \cite{5}, the maximum of the ion distribution function over
the energy is at a level of 0.8--0.85 times the applied voltage. In our
case, the typical discharge voltage varied from 3.5 to 4~kV, so that the
energy of ions in the distribution function maximum equals 2.8--3.4~keV.
Therefore, in subsequent calculations, we will use this energy range.

The surface of a solid is known to often have rather a developed relief
structure. It is also known that the interaction with ionic and plasma fluxes
induces those or other relief modifications. Depending on the flux
parameters and the conditions at the surface, these modifications manifest
themselves as both the development and the smoothing of a relief. The
effect of smoothing is usually observed either at ion energies below the
sputtering threshold or at large incidence angles, when the sputtering
coefficient is lower than that at normal incidence \cite{17}. In the case of
a reflex discharge, since plasma rotates, the angle of ion incidence onto the
cathode surface can differ considerably from zero, which leads, in turn, to a
modification of the sputtering coefficient. The dependence of the sputtering
coefficient $Y$ on the ion incidence angle $\theta $ is expressed by the
formula \cite{19}
\begin{equation}
Y(\theta )=Y(0)x^{f}\exp\left[ -\Sigma (x-1)\right] ,
\end{equation}
where $x=1/\cos \theta $, and $f$ and $\Sigma $ are parameters that
are either determined from the experiment or calculated. The results
of corresponding calculations by formula~(2) are shown in Fig.~3.
The figure demonstrates that the angle, at which the maximum of the
sputtering coefficient is observed, is equal to about 73$^{\circ }$,
and the sputtering coefficient at the maximum is approximately 3
times higher.

\begin{figure}
\includegraphics[width=8.2cm]{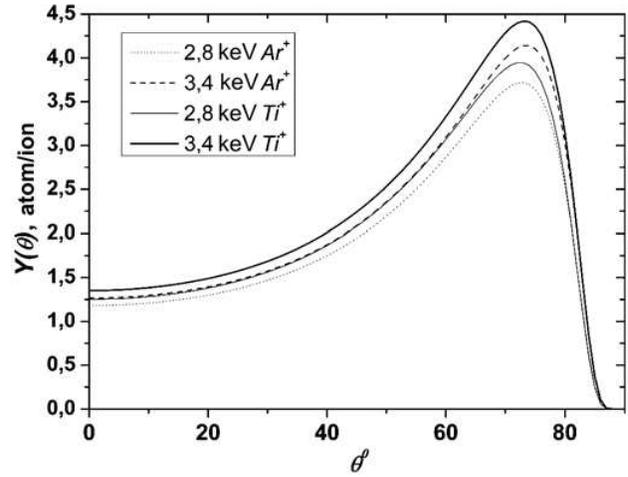}
\vskip-3mm\caption{Dependences of the sputtering coefficient on the
angle of ion incidence onto the target for various ion energies  }
\end{figure}

The total number of particles sputtered from the cathode surface within the
time interval $t$ equals
\begin{equation}
N_{\Sigma m}=\Gamma _{i}YS_{\Sigma }t,
\end{equation}
where $S_{\Sigma }=0.01571~\mathrm{m}^{\mathrm{2}}$ is the total area of
cathodes, $\Gamma _{i}$ is the particle flux onto the cathode surface
[m$^{-2}\mathrm{s}^{-1}$], and $t$ is the time [s]. The quantity $\Gamma _{i}$
is determined as follows:
\begin{equation}
\Gamma _{i}=N_{i}v_{s},
\end{equation}%
where $N_{i}$ is the ion concentration [m$^{-3}$], and $v_{s}$ the
ion-sound velocity (the velocity of ion sound) determined as
\begin{equation}
v_{s}=9.79\times 10^{3}\left( ZT_{e}/M_{i}\right) ^{1/2},
\end{equation}
where $T_{e}$ is the electron temperature [eV], $M_{i}$ the ion mass
[a.m.u.], and $Z$ the ion charge. We adopt that the ion mass equals the mass
of an argon atom, $M_{i}=39.94$, and the electron temperature $T_{e}=(1\div
10)$~eV to obtain $v_{s}=(1.5\div 4.9)\times 10^{3}$~m/s.

As was indicated earlier in work \cite{14}, the time dependence of the average
concentration in a gas-metal plasma can be divided into three stages:
the formation, existence, and decay of a dense plasma. To estimate the total
number of sputtered particles at the plasma formation stage, $N_{\Sigma m}$,
we accept the variation of the particle concentration in time to be equal to that
experimentally obtained in work \cite{14}, the time
$t\approx100$~$\mu\mathrm{s}$, and the ion energy equal 0.8--0.85 times the applied voltage.
In this case, the magnitude of $N_{\Sigma m}$ ranges from $7.5\times10^{16}$
to $2.8\times 10^{17}$~particles, depending on the sputtering coefficient
value. At the stage of dense plasma existence, the average sputtering
coefficient is adopted to equal 0.02--0.26, and the time
$t\approx800$~$\mu\mathrm{s}$, so that we obtain $N_{\Sigma
m}\sim10^{17}\div10^{18}~$particles at this stage of the discharge, which is in agreement with
experimental and theoretical results of work \cite{15}, $N_{\Sigma
m}=9.26\times10^{16}\div8.5\times10^{17}~$particles.

\begin{figure}
\includegraphics[width=\column]{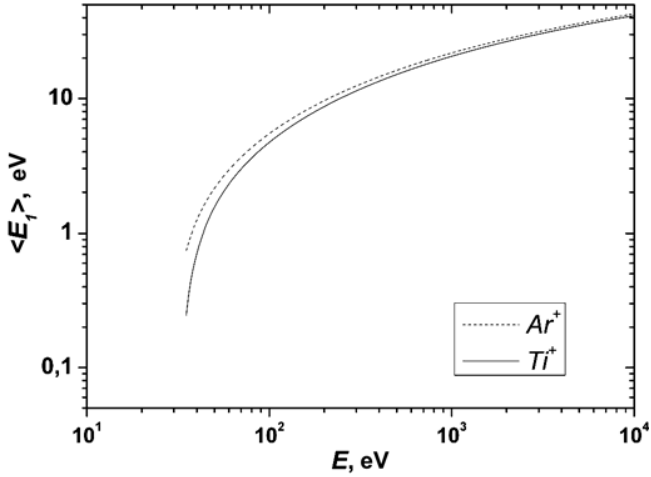}
\vskip-3mm\caption{Average energy of titanium atoms escaping from
the target as a function of the incident ion energy  }
\end{figure}

\section{Efficiency of Ionization Processes of Sputtered Atoms and Formation
of Gas-Metal Plasma}

The efficiencies of the processes of sputtered atom capture into the discharge
and gas-metal plasma formation depend on two processes, namely, the diffusion
and the ionization of atoms in the primary plasma. Let us consider these processes
in more details.

According to the kinetic theory of gases, the diffusion coefficient equals
\cite{20}
\begin{equation}
D=\frac{1}{3}\lambda v,
\end{equation}
where $\lambda $ is the mean free path [m], and $v$ is the velocity
[m/s]. The characteristic diffusion time is equal to the time of
particle arrival at the wall,
\begin{equation} \tau
_{D}=\frac{\Lambda ^{2}}{D},
\end{equation}
where $\Lambda $ is the characteristic diffusion length, which, in the case
of cylindrical geometry, satisfies the relation
\begin{equation}
\frac{1}{\Lambda ^{2}}=\left( \frac{2.4}{R}\right) ^{2}+\left( \frac{\pi
}{l}\right) ^{2},
\end{equation}%
where $R$ is the radius of the system [m], and $l$ is its length [m]. In
order to determine the mean free path from the energy of a sputtered atom
moving in the gas, the atoms of which have the Maxwellian distribution over
their velocities, one may use the expression obtained in work \cite{21},
\begin{equation}
\lambda =\lambda _{0}\left[ \left( 1+\frac{1}{2\omega }\right) {\rm
erf}\left( \sqrt{\omega }\right) +\frac{e^{-\omega }}{\sqrt{\pi
\omega }}\right] ^{-1},
\end{equation}
\begin{equation}
\omega =\frac{3}{2}\frac{E_{1}}{E_{g}}\frac{M_{g}}{M_{m}},
\end{equation}%

\noindent where $E_{1}$ is the energy of a sputtered atom
[eV];\thinspace $E_{g}$ is the average energy of a gas atom [eV];
$M_{g}$ and $M_{m}$ are the masses of gas and sputtered atoms,
respectively [a.m.u.]; $\lambda _{0}=1/N\sigma $; $\sigma $ is the
effective collision cross-section [m$^{2}$]; and $N$ is the gas
particle concentration [m$^{-3}$]. When a sputtered atom moves in
the gas, its energy relaxes owing to collisions, and the average
energy of atoms, $E_{F}$, at some distance from the sputtered
surface can be estimated as \cite{22}
\begin{equation}
E_{F}=\left( E_{0}-kT_{g}\right) \exp\left[ n\ln \left(
\frac{E_{f}}{E_{i}}\right) \right] +kT_{g},
\end{equation}%
where $E_{0}$ is the initial energy of a sputtered atom [J]; $k$ is
the Boltzmann constant [J/K]; $T_{g}$ is the gas temperature [K];
$E_{f}/E_{i}=\Delta E/E=2M_{g}M_{m}/(M_{g}+M_{m})^{2}$ is the ratio
between the energies after and before the collision; $n$ is the
number of collisions, which is determined by the formula
$n=l_{1}P\sigma /kT_{g}$, where $l_{1}$ is the path length [m]; and
$P$ is the gas pressure [Pa]. For the estimation of $E_{F}$, we
adopt that $E_{0}=\left\langle E_{1}\right\rangle $ and
$T_{g}=300~\mathrm{K}$.

The average energy of atoms that escape from the target is determined as
\cite{23}
\begin{equation}
\left\langle E_{1}\right\rangle =2U_{s}g(w),
\end{equation}
where the function $g(w)$ looks like
\begin{equation}
g(w)=\left( \ln (w)+\frac{2}{w}-\frac{1}{2w^{2}}-\frac{3}{2}\right) \left(
1-\frac{1}{w}\right) ^{-2},
\end{equation}%
and $w=E/E_{\mathrm{th}}$. The results of calculations carried out for the
average energy of titanium atoms are depicted in Fig.~4. The calculated
dependences of the quantity $E_{F}$ on the gas pressure are shown in Fig.~5.
It is worth noting that a reduction in the energy of sputtered atoms in
the course of their motion in the gas (see Fig.~5) results in a decrease
of the atomic path length; however, a considerable influence takes place, when
the energy of sputtered atoms approaches the thermal one. Despite that the
transport scattering cross-section is larger than the gas-kinetic one by not
less than an order of magnitude, the collisions of sputtered atoms with gas ions
prevail only at high gas ionization degrees.

The energy spectrum of sputtered atoms is described by Thompson's formula
\cite{24}. With regard for the anisotropic effects and the angle of ion
incidence, it reads \cite{25}
\begin{equation}
\Phi \left( E_{1},\theta \right) \propto \frac{E_{1}\cos \theta }{\left(
E_{1}+U_{s}\right) ^{4}}\left( E_{1}\cos ^{2}\theta +U_{s}\right).
\end{equation}%

\begin{figure}
\includegraphics[width=8cm]{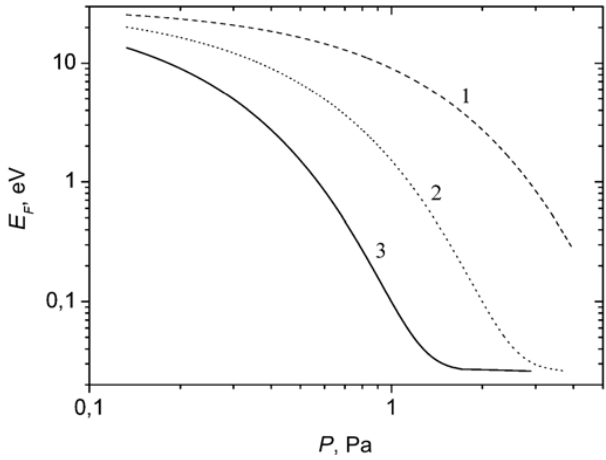}
\vskip-3mm\caption{Dependences of the average energy of sputtered
titanium atoms on the argon pressure in a discharge chamber for
various distances from the sputtered surface: 0.02 (\textit{1}),
0.05 (\textit{2}), and 0.1~m~(\textit{3})  }
\end{figure}

\noindent Using Eqs.~(6)--(10) and averaging over the distribution
function of sputtered atoms over the energy, which depends on the
ion incidence angle, we obtain the average value for the time of
sputtered atom diffusion to the chamber wall. The result of
calculations is depicted in Fig.~6 (curves \textit{6} and
\textit{7}).

Electron impact ionization is one of the main processes that give rise to
the atomic ionization. In this case, the ionization time equals
\begin{equation}
\tau _{i}=\frac{1}{N_{e}\left\langle \sigma _{e}v_{e}\right\rangle
},
\end{equation}%
where $\left\langle \sigma _{e}v_{e}\right\rangle $ is the rate of
atom ionization by the electron impact [m$^{3}/\mathrm{s}$]
\cite{26}, and $N_{e}$ is the electron concentration [m$^{-3}$]. An
additional mechanism of ionization for titanium atoms can be the
following processes \cite{27}: (i)~an ion charge exchange at an atom
(the non-resonance charge exchange), $X^{+}+Y\rightarrow
X+Y^{+}+\Delta E$ or $X^{+}+Y\rightarrow X+Y^{+\ast }+\Delta E$,
where $\Delta E$ is the mismatch of the process energy, which is equal
to the difference between the ionization or excitation potentials of
both colliding particles; (ii)~ionization at the collision with a
metastable atom (the Penning process), $X^{\ast }+Y\rightarrow
X+Y^{+}+e$. According to work \cite{28}, the rate of charge exchange
for an argon ion with a titanium one is $k_{\rm CT}=6.61\times
10^{-15}~\mathrm{m}^{3}/\mathrm{s}$, and the rate of Penning process
is $k_{\rm PI}=2.75\times 10^{-16}~\mathrm{m}^{3}/\mathrm{s}$.
Therefore, the total ionization time, taking the additional
processes into account, equals
\begin{equation}
\tau _{i}=\frac{1}{N_{e}\left\langle \sigma _{e}v_{e}\right\rangle
+N_{i}k_{\rm CT}+N_{m}k_{\rm PI}}.
\end{equation}

\begin{figure}
\includegraphics[width=\column]{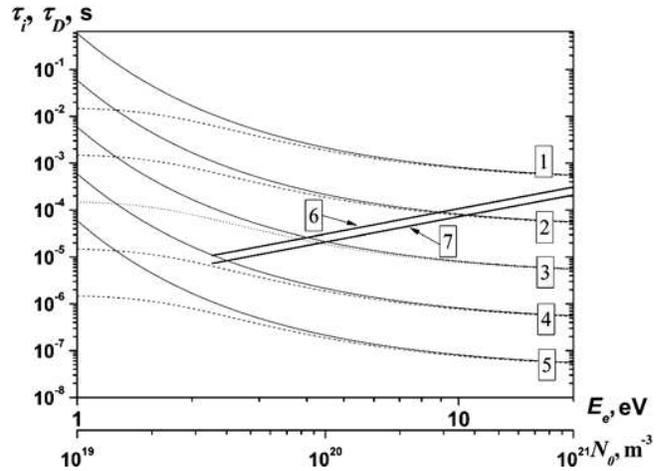}
\vskip-3mm\caption{Dependences of the titanium atom ionization time on
the electron energy for various electron concentrations: $10^{16}$
(\textit{1}), $10^{17}$ (\textit{2}), $10^{18}$ (\textit{3}),
$10^{19}$ (1), and $10^{20}$~\textrm{m}$^{-3}$ (\textit{5}); and the
dependences of the titanium atom diffusion time on the initial
concentration of neutral particles of a working substance in the gas
phase for various incidence angles $\theta=73$ (\textit{6}) and
0$^{\circ}$ (\textit{7}). The dash-dotted and solid curves were
calculated taking and not taking the charge exchange into account,
respectively }
\end{figure}

As was shown in work \cite{29}, the Penning ionization dominates over the electron
impact ionization in the concentration range below $2\times
10^{16}$~\textrm{m}$^{-3}$ and at low temperatures. Therefore, at higher concentrations, it
can be neglected. In Fig.~6, the dependences of the titanium atom ionization
time are shown for various initial electron concentrations; the curves were
calculated taking the non-resonance charge exchange into account and without
this account (i.e. accepting $N_{e}=N_{i}$). One can see that the
contribution made by the charge exchange becomes substantial only if $T_{e}\leq 3
$~eV. A comparison between the characteristic times $\tau _{i}$ and $\tau
_{D}$ (see Fig.~6) gives only a qualitative picture of the influence exerted
by the processes of ionization and diffusion of sputtered atoms. However, it
does not allow one to evaluate the ionization degree of sputtered atoms
quantitatively. In the stationary case where the ionization by electrons is the
major ionization process, the balance equations for particles in plasma can
be written down in the form
\[
\langle\sigma_e v_e \rangle N_e
N_{\mathrm{Ti}}=\frac{N_{\mathrm{Ti}^+}}{\tau_{\mathrm{Ti}^+}},
\]
\[
\langle\sigma_e v_e \rangle N_e
N_{\mathrm{Ar}}=\frac{N_{\mathrm{Ar}^+}}{\tau_{\mathrm{Ar}^+}},
\]
\begin{equation}
N_e=N_{\mathrm{Ar}^+}+N_{\mathrm{Ti}^+},
\end{equation}

\begin{figure}
\includegraphics[width=7.8cm]{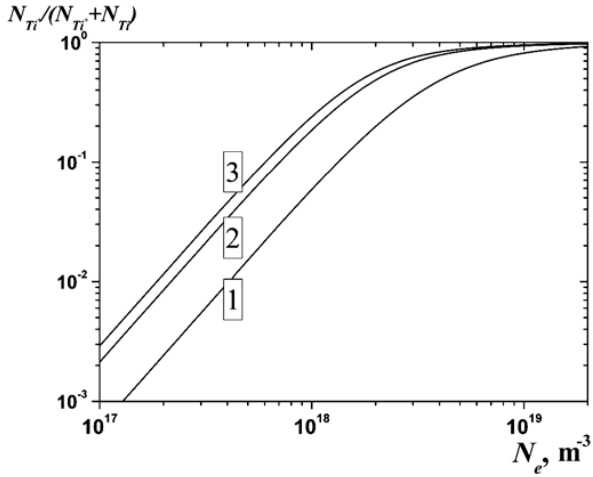}
\vskip-3mm\caption{Dependences of the titanium atom ionization degree on
the electron concentration at various $T_{e}=3$ (\textit{1}), 6
(\textit{2}), and 8~eV (\textit{3})  }
\end{figure}

\noindent where $N_{\mathrm{Ti}}$ and $N_{\mathrm{Ar}}$ are the
concentrations of neutral titanium and argon atoms, respectively;
$N_{\mathrm{Ti}^{+}}$ and $N_{\mathrm{Ar}^{+}}$ are the
concentrations of titanium and argon ions, respectively; and $\tau
_{\mathrm{Ti}^{+}}$ and $\tau _{\mathrm{Ar}^{+}}$ are the
corresponding lifetimes of ions in plasma. Taking Eq.~(17) into
account, the ionization degree can be expressed as follows:
\begin{equation}
\frac{N_{\mathrm{Ti}^{+}}}{N_{\mathrm{Ti}}+N_{\mathrm{Ti}^{+}}}=\frac{\left\langle
\sigma _{e}v_{e}\right\rangle N_{e}\tau
_{\mathrm{Ti}^{+}}}{1+\left\langle \sigma _{e}v_{e}\right\rangle
N_{e}\tau _{\mathrm{Ti}^{+}}}.
\end{equation}

For the case of a pulsed discharge where the characteristic times of the plasma
formation are shorter than the time of the ionization equilibrium establishment,
the ionization degree strongly depends on the existence time of plasma with
a given concentration. In our case, the characteristic times for
concentrations of $1.4\times 10^{18}$ and $2\times 10^{19}$~\textrm{m}$^{-3}$
to be attained are equal to 15 and 100~$\mu \mathrm{s}$, respectively
\cite{14}. In the general case, in order to find the ionization degree, it is
necessary to solve a system of differential equations. At the initial stage
of plasma generation, the ion lifetime is usually larger than the time of
the concentration growth, i.e. $\tau >t$. In view of this inequality,
the system of equations looks like
\begin{equation} \left\{ \begin{array}{l} \frac{dN_{e}}{dt}=\left\langle \sigma
_{e}v_{e}\right\rangle N_{e}N_{\mathrm{Ar}}+\left\langle \sigma
_{e}v_{e}\right\rangle N_{e}N_{\mathrm{Ti}},
  \\ [3mm]
\frac{dN_{\rm Ar}}{dt}=-\left\langle \sigma _{e}v_{e}\right\rangle
N_{e}N_{\mathrm{Ar}}+N_{\mathrm{Ar}^{+}}N_{\mathrm{Ti}}k_{\rm
CT},   \\
[3mm]\frac{dN_{\mathrm{Ar}^{+}}}{dt}=-\left\langle \sigma
_{e}v_{e}\right\rangle
N_{e}N_{\mathrm{Ar}}-N_{\mathrm{Ar}^{+}}N_{\mathrm{Ti}}k_{\rm CT},
 \\
[3mm]\frac{dN_{\mathrm{Ti}}}{dt}=\frac{\Gamma S_{\Sigma
}Y}{V}-\frac{N_{\mathrm{Ti}}}{\tau _{D}}-\left\langle \sigma
_{e}v_{e}\right\rangle
N_{e}N_{\mathrm{Ti}}-N_{\mathrm{Ar}^{+}}N_{\mathrm{Ti}}k_{\rm CT},
\\ [3mm] \frac{dN_{\mathrm{Ti}^{+}}}{dt}=\left\langle \sigma
_{e}v_{e}\right\rangle
N_{e}N_{\mathrm{Ti}}+N_{\mathrm{Ar}^{+}}N_{\mathrm{Ti}}k_{\rm CT}.
\end{array}\right.
\end{equation}

To solve the differential equations numerically, let us set the initial
conditions in accordance with experimental data \cite{14}. Namely, the
discharge voltage is 3.5~kV, i.e. $E\approx 2.8$~keV; the sputtering
coefficients are taken in accordance with Figs.~2 and 3; the diffusion time
of titanium atoms is taken according to Fig.~6; the variation of the electron
concentration corresponds to the experiment; $N_{\mathrm{Ar}}=7\times 10^{19}
$~\textrm{m}$^{-3}$; and, since $N_{\mathrm{Ar}}>N_{\mathrm{Ti}}$, the
second equation in system~(19) is ignored. The results of calculations for
the dependences of the ionization degree of titanium atoms on the electron
concentration and the temperature are presented in Fig.~7 (the electron
distribution function over the energy was assumed to be Maxwellian). One can
see that, at a concentration of $2\times 10^{19}$~\textrm{m}$^{-3}$
($t\approx 100$~$\mu \mathrm{s}$), the ionization degree is close to 100\%.
The content of titanium ions averaged over the volume varies from 10 to
40\%, depending on the sputtering coefficient.

As is seen from Figs.~6 and 7, the efficiency of the gas-metal plasma
formation (or, equivalently, the introduction of a working metal
substance into plasma followed by the metal atom ionization and
the gas-metal plasma formation) governed by the mechanism of cathode
material sputtering substantially depends on the initial plasma
concentration. The highest efficiency of the gas-metal plasma generation
is observed for the electron concentration $N_{e}\geq
10^{19}$~\textrm{m}$^{-3}$ (see Figs.~6 and 7). In spite of the fact
that our calculations were carried out for one cathode material --
namely, titanium -- a similar picture should be observed for other
metals as well, because their ionization potentials and ionization
cross-sections are close to one another. This
hypothesis is confirmed by the results obtained on an \textquotedblleft
ERIC\textquotedblright\ installation \cite{12}, where the ignition
gases Ar and Kr were used. When Ni, Cu, Pd, and Gd atoms were
introduced at $N_{e}=4\times 10^{16}\div 2\times
10^{17}$~\textrm{m}$^{-3}$ and $T_{e}=3\div 6~\mathrm{eV}$, the
ionization probability for sputtered atoms amounted to 4--14\%, and
the relative concentration of metal ions in plasma was 3--17\%.

The proposed model can be applied at the initial stage of plasma
formation, when the ion lifetime is longer than the time of
concentration growth, i.e. $\tau >t$. In the other case, e.g., under
stationary conditions, it is necessary that the ion lifetime should
be taken into account (see Eqs.~(17) and (18)); it consists of the
recombination and diffusion times. The model does not consider
the dependence of the sputtering coefficient on the target
temperature, which considerably increases at $T>0.7T_{m}$, where
$T_{m}$ is the melting temperature \cite{17}. It also does not involve
the cluster formation from two and more atoms, when the
target is sputtered. In addition, there is an uncertainty in the angle
of ion incidence onto the target. On the one hand, this fact is
associated with initial conditions at the target surface (the
surface relief). On the other hand, the angle of incidence is
determined by the ion motion in crossed electric and magnetic fields
before the ion collides with the target, so that the incidence angle
can differ considerably from zero. However, the results obtained are
in satisfactory agreement with experimental data \cite{13,14,15}.
Nevertheless, the estimations made with the use of the given model cannot be
considered complete and ultimate. They demand that the model should
be more specified according to the variation of experimental
conditions.\looseness=1

\section{Conclusions}

1. According to the stated objective, a model for the quantitative
estimation of the entry of heavy sputtered components into the discharge is
selected. In the framework of the model used in this work, the evaluation of
the parameters of a sputtering mechanism, owing to which the working substance
penetrates into the discharge, is carried out. The influence
of the incident ion energy is considered for the $\mathrm{Ar}^{+}\rightarrow
\mathrm{Ti}$ and $\mathrm{Ti}^{+}\rightarrow \mathrm{Ti}$ ion-atom pairs in the
cases of normal incidence and when the angle of ion incidence onto the
target varies. The values of sputtering coefficient calculated numerically
are in satisfactory agreement with experimental data obtained under similar
conditions and published earlier in work \cite{15}. The most rapid variation
of the sputtering coefficient occurs at the energy of incident ions lower than
or equal to 200~eV and at incidence angles of 40--85$^{\circ }$.

\noindent 2. To determine the efficiency of the gas-metal plasma
formation (or, equivalently, the ionization of a gas-metal mixture),
the following dependences are calculated: the dependences of the
average energy of titanium atoms that escape from the target on the
energy of incident argon and titanium ions and the pressure of argon
in a discharge chamber; the dependence of the ionization time of
titanium atoms on the electron energy; the dependence of the atom
diffusion time on the initial concentration of neutral particles;
and the dependence of the ionization degree of titanium particles on the
electron concentration. The most effective generation of a gas-metal
plasma is observed provided the electron concentration $N_{e}\geq
10^{19}$~\textrm{m}$^{-3}$. The results obtained agree well with
experimental ones \cite{13,14,15}. At the same time, the applied
model should be additionally specified in the case where
experimental conditions vary.\looseness=1

\rezume{%
ОЦІНЮВАННЯ ЕФЕКТИВНОСТІ\\ ВВЕДЕННЯ РЕЧОВИНИ У ВІДБИВНИЙ РОЗРЯД\\ ЗА
РАХУНОК РОЗПИЛЮВАННЯ МАТЕРІАЛУ КАТОДА}{Ю.В. Ковтун, Є.І. Скібенко,
А.І. Скибенко,  В.Б. Юферов} {У роботі розглянуто процеси, пов'язані
з введенням  робочої речовини у плазму імпульсного відбивного
розряду за рахунок розпилюючого механізму з метою створення густої
багатокомпонентної газометалевої плазми. При цьому запропоновано
розрахункову  модель оцінки параметрів розпилюючого механізму, за
рахунок якого потрібна робоча речовина надходить у розряд. Одержані
дані задовільно узгоджуються з результатами експерименту.}

\end{document}